\documentstyle[11pt,newpasp,twoside,epsf]{article}
\begin{document}

\title{Radio and gamma-ray emission from pulsars}

\author{Qiao, G. J., Lee, K. J., Wang, H. G., and Xu, R. X.}

\affil {Astronomy Department, Peking University,
       Beijing 100871, P.R.China. }


\begin{abstract}

The radiation of pulsars have been observed for many years.
A few pulsars are discovered to have both radio and gamma-ray
emission. Many models on pulsar radiation have been developed, but
so far we are still lacking an elaborate model which can explain
the emission from radio to gamma-rays in detail. In this paper we
present a joint model for radio and gamma-ray emission, in which
both the dominate emission mechanisms are inverse Compton
scattering. The pulse profiles at radio and gamma-ray bands are
reproduced for the Crab-like, Vela-like and Geminga-like pulsars,
in good agreement with observations.
\end{abstract}


\section{Introduction}

A wealth of observational data on radio pulsars has accumulated in
the last three decades. So far, seven gamma-ray pulsars have been
discovered, of which six are also radio pulsars. The general
properties of gamma-ray pulsars could be summarized as follows
(Kanbach 2002): (1) the light curves of most gamma-ray pulsars
show two major peaks with pulse widths close to 50 percent of the
rotation period, (2) the gamma-ray spectra are hard, normally with
a luminosity maximum around 1 GeV, but a spectral cutoff above
several GeV is found; (3) the spectra vary with rotational phase
indicating different sites of emission; (4) the gamma-luminosity
scales with the particle flux from the open regions of the
magnetosphere (Goldreich-Julian current). Except in the case of
the Crab pulsar, the gamma-ray pulses are mostly much wider than
the radio mean pulses and often show remarkable phase offsets from
the radio pulses (e.g., the Vela pulsar).

In history, models for pulsar radio and gamma-ray emission have
been developed within two separated domains, namely, the polar cap
model for radio emission, either the polar cap or the outer gap
models for gamma-ray emission. Since both radio and gamma-ray
emissions can be observed simultaneously from a same pulsar, there
come the questions: is there any relationship between radio and
gamma-ray radiation? Is there any interaction between the
acceleration processes in the polar and outer-gap accelerators?
Can we have a model for both radio and gamma-ray emission?

There have been some efforts to fit the observations. It is
suggested that in the outer gap model the gamma-ray emission
altitude should be lower (Cheng \& Zhang 1999), whereas in the
polar cap model the altitude should be higher (Harding \& Muslimov
1998), in order to explain well the observations. However, the
phase shifts of Crab-like and Vela-like pulsars, between radio and
gamma-ray pulses, can not be explained simultaneously with either
the outer or the polar cap models. If the emission is located
between the null surface and the polar cap, a lot of observational
facts (including the relative phase shifts) may be understood.
Null surface plays a vital role in such a scenario.

This paper is an effort of such a joint model. Radio emission
altitudes and pulse profiles are figured out under the ICS model
(e.g., Qiao \& Lin 1998), gamma-ray pulse profiles are calculated
by assuming a second acceleration region above the polar gap (that
is produced by space-charge limited flow of secondary particles).

\section{A short statement for radio emission}

We have proposed an inverse Compton scattering (ICS) model for
pulsar radio emission (Qiao \& Lin 1998, Xu et al. 2000, Xu et al.
2001, Qiao et al. 2001). Some main observational features can be
understood under this model: (1) the central or `core' emission
beam and the conal beams; (2) emission altitude for each emission
component; (3) linear and circular polarization of individual and
integrated pulses; (4) pulse profiles changing with frequency. The
equations what we used in this paper can be found in the papers
above.

\section{Equations for gamma-ray emission}
Owing to there is a strong acceleration process through space
charge limited flow near null surface, Gamma-rays would be
produced there by synchrotron and Inverse Compton scattering
mechanisms.

The
emission location with distance $r=\lambda R_{\rm null}$ from the
stellar center, where $R_{\rm null}$ is the minimum distance from
the center to the intersection of null charge surface and the last
open field lines, $\lambda$ is a constant which is supposed to be
less than but close to 1 (so that the acceleration region is not
far from the null surface), e.g., 0.8 in this paper. Here we only
show an analytical approximation for a small inclination angle
($\alpha$). For a large $\alpha$® we perform numerical methods.

The emission location could be determined by
\begin{equation}
\theta_{\rm e}=\arcsin (\sqrt{\lambda}\sin(\frac{1}{2}(\arccos(-
\frac{\cos\alpha}{3})-\alpha))),
\end{equation}
where $\theta_{\rm e}$ is the angle between the magnetic axis and
position vector ${\vec r}$, and $\alpha$ is the inclination angle.
The angle between the radiation direction and the magnetic axis,
$\theta _{\rm \mu}$, is given by
\begin{equation}
\theta_{\rm \mu}=-\alpha-\arctan(\frac{\cos(\alpha)+3
\cos(\alpha+2 \theta_{e})}{\sin(\alpha)
+3\sin(\alpha+2\theta_{e})})+\frac{\pi}{2}.
\end{equation}
With the emission angle we can calculate the phase separation
between two emission components. Supposing that each emission
component is Gaussian, we fit the mean pulse profiles with
observations (see next section for results).

\section{Results}

The observed pulse profiles in gamma-ray and radio bands are
fitted for seven pulsars, i.e., the Crab, PSR B1509-58, the Vela,
PSR B1706-44, PSR B1951+32, the Geminga and PSR B1055-52. The
profiles and fitted curves are shown in Figure 1. It is found that
the observational profiles can be reproduced when reasonable
parameters are chosen in our simulations.

\section{Conclusions and discussions}

The pulse profiles at both radio and gamma-ray bands are
calculated under the joint model as it is shown here. In
principle, the main observational features can be reproduced with
this model. More details, such as phase resolved spectrum at
gamma-rays, X-ray emission from pulsars, and so on, are need to be
investigated in the future. For the inner gap sparking, a strange
star model is benefit(Xu et al.2001).

\begin{acknowledgements}
We are very grateful to Prof. R. N. Manchester for his valuable
discussions. This work is supported by NSF of China, the Research
Fund for the Doctoral Program Higher Education, and the Special
Funds for Major State Basic Research Projects of China
(G2000077602).
\end{acknowledgements}

\clearpage

\begin {figure}
\plotone{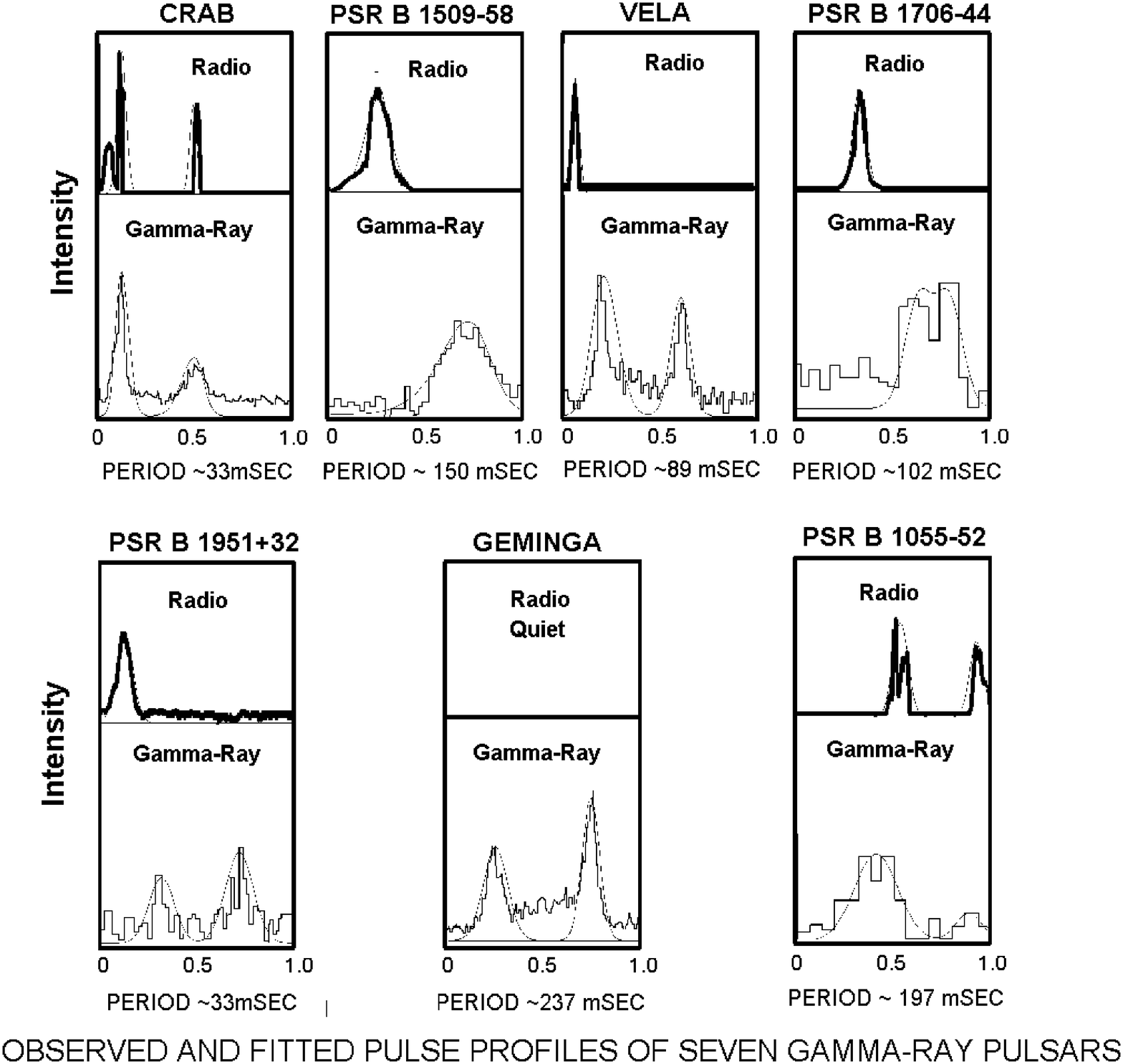} \caption{The observed and the simulated mean
pulse profiles for seven pulsars. The dash lines are for
simulations, and the solid lines for the observations.}
\end {figure}

\end{document}